\documentclass[a4paper]{jpconf}
\usepackage{graphicx}
\begin{document}
\title{Glow-in-the-dark globular clusters: modelling their multiwavelength lanterns}

\author{C Venter$^1$, I Buesching$^1$, A Kopp$^2$, A-C Clapson$^3$, and O C de Jager$^1$}

\address{$^1$Centre for Space Research, North-West University, Potchefstroom Campus, Private Bag X6001, Potchefstroom 2520, South Africa\\
$^2$Institut f\"{u}r Experimentelle und Angewandte Physik, Christian-Albrechts-Universit\"{a}t zu Kiel, Leibnizstrasse 11, 24118 Kiel, Germany\\
$^3$Max-Planck-Institut f\"{u}r Kernphysik, PO Box 103980, 69029 Heidelberg, Germany}

%\ead{jacky.mucklow@iop.org}

\begin{abstract}
Globular clusters (GCs) are astronomical tapestries embroidered with an abundance of exotic stellar-type objects. %, including ancient metal-poor stars, planetary nebulae, white dwarfs (WDs), low-mass X-ray binaries (LMXRBs), RR Lyrae variables, blue stragglers, cataclysmic variables, and possibly even central black holes. In addition, 
Their high age promises a rich harvest of evolved stellar products, while the deep potential wells and high mass densities at their centres probably facilitate the formation of multiple-member stellar systems via increased stellar encounter rates. The ubiquity of low-mass X-ray binaries, thought to be the progenitors of millisecond pulsars (MSPs), explain the large number of observed GC radio pulsars and X-ray counterparts. %that have already been discovered. 
The \textit{Fermi} Large Area Telescope (LAT) recently unveiled the first $\gamma$-ray GC pulsar (PSR~J1823$-$3021A). The first observations of GCs in the GeV and TeV bands furthermore created much excitement, and in view of the above, it seems natural to explain these high-energy lanterns by investigating an MSP origin. An MSP population is expected to radiate several pulsed spectral components in the radio through $\gamma$-ray wavebands, in addition to being sources of relativistic particles. The latter may interact with background photons in the clusters producing TeV excesses, while they may also radiate synchrotron photons as they traverse the cluster magnetic field. We present multiwavelength modelling results for Terzan~5. %, focusing on the system constraints that may be derived in the context of this model by comparing our model to multiwavelength data. 
We also briefly discuss some alternative interpretations for the observed GC $\gamma$-ray signals.
\end{abstract}

\section{Introduction}
There are about 150 Galactic globular clusters (GCs), ancient spherical arrangements of $10^5-10^6$ stars bound by their mutual gravity. The peculiar properties of these objects have been useful in diverse astrophysical disciplines such as cosmology, galaxy formation, stellar evolution and dynamics, and binary and variable stars~\cite{Harris96}. Empirical correlations between (i) the number of low-mass X-ray binaries (LMXRBs) -- the progenitors of millisecond pulsars (MSPs) -- and the two-body stellar encounter rate~\cite{Pooley03}, and between (ii) the number of MSPs ($N_{\rm MSP}$) and this same encounter rate~\cite{Abdo10,Hui10}, provide evidence for a dynamical formation of such GC stellar objects.
Encounter rates may in turn be enhanced by the high core densities of GCs~\cite{Verbunt87}, leading to relatively large numbers of LMXRBs and MSPs in GCs. These sources may be directly or indirectly responsible for, or contribute to, the multiwavelength emission seen from GCs. This paper discusses the multiwavelength observations and modelling of GCs, but with a strong focus on the $\gamma$-ray waveband.
%HE $L_\gamma$ and $u$ \cite{Hui10}
%http://www.naic.edu/~pfreire/GCpsr.html

\section{Multiwavelength pieces of the puzzle}
GCs have now plausibly been detected in all energy domains~\cite{Abdo09_Tuc}. There are various examples: surface brightness profile measurements in the optical and infrared which facilitate King-model-type fits~\cite{Trager95} that in turn allow, e.g., inference of structural parameters; discoveries of embedded radio MSPs~\cite{Ransom08}, X-ray MSPs, LMXRBs, white dwarfs (WDs), and main-sequence binaries~\cite{Grindlay05}; extended radio emission~\cite{Clapson11}; and the high-energy (HE) and very-high-energy (VHE) detections discussed below.

\subsection{GCs by the dozen: \textit{Fermi} Large Area Telescope (LAT) results}
\label{sec:Fermi}
Just over a year into the \textit{Fermi} mission, detection of 47~Tucanae at a $17\sigma$ level was claimed, making this the first GC to be detected in $\gamma$-rays~\cite{Abdo09_Tuc}. The spectrum was consistent with being an accumulation of several individual MSPs' pulsed spectra (for an alternative interpretation, see Section~\ref{sec:Cheng}), with the measured luminosity constraining the 47~Tucanae MSP population to $\sim60$. However, no pulsations from individual MSPs were found. This was followed by the detection of Terzan~5~\cite{Kong10}, as well as several (plausibly $\sim6$) other GCs~\cite{Abdo10}, including M~28, NGC~6388, and Omega Cen. Another 3 detections and 3 GC candidates were claimed~\cite{Tam11}, bringing the total number of HE GCs to about a dozen~\cite{2FGL}. 

\subsection{The shy waveband: VHE observations}
In contrast to the HE successes, TeV observations have only been able to uncover an excess in the direction of, but offset from the centre of, Terzan~5~\cite{Abramowski11_Ter5} (with a probability of a chance coincidence of $\sim10^{-4}$). The extended VHE source reaches beyond the cluster's tidal radius, exhibiting a power-law spectrum with a photon index of $2.5\pm0.3_{\rm stat}\pm 0.2_{\rm sys}$ that implies an integral flux of $(1.2 \pm 0.3) \times 10^{-12}$~cm$^{-2}$\,s$^{-1}$ above 440~GeV. Only upper limits exist on other GC positions: 47 Tuc~\cite{Aharonian09_Tuc}, M~13~\cite{Anderhub09,McCutcheon09}, M~5 and M~15~\cite{Abramowski11_NGC6388,McCutcheon09}, and NGC~6388~\cite{Abramowski11_NGC6388}.

\subsection{Diffuse X-rays}
Diffuse X-rays have been reported for M~22, Omega Cen, 47~Tucanae~\cite{Hartwick82,KG95}, M~80, NGC~6266, NGC~6752, and M~5~\cite{Ok07}. This emission has been interpreted~\cite{KG95} as being due to the formation of bow shocks which result from the GC motion through the Galactic halo plasma, since it appears to be associated with GCs with high proper motions and large accumulations of internal gas~\cite{Ok07}. However, the clumpy structure observed near NGC~6752 exhibits a hard non-thermal spectrum as well as a radio counterpart, and may be due to bremsstrahlung by shock-accelerated electrons hitting nearby gas clouds, while unresolved X-ray sources may contribute to, or even dominate, diffuse X-ray radiation from GCs in general~\cite{Eger10}. 
Diffuse X-ray emission has also been observed from Terzan~5 (1--7~keV), peaking at the centre and decreasing smoothly outwards~\cite{Eger10}, probably of a non-thermal origin (e.g., synchrotron radiation -- SR). A follow-up search for diffuse X-rays from six HE GCs yielded no significant emission above the background level~\cite{Eger12}. Within the MSP scenario that this paper focuses on, SR is expected to be produced by relativistic leptons that escape from a population of MSPs inside GCs and interact with the GC magnetic field~\cite{Venter08_conf}. Using the diffuse X-ray profile~\cite{Eger10}, the diffusion coefficient of these particles may be constrained~\cite{Cheng10,Kopp13}. One can also probe the magnitude and profile of the cluster's magnetic field.

\section{Message from the most powerful $\gamma$-ray MSP}
\label{sec:1823}
The \textit{Fermi} detection~\cite{Freire11} of the most powerful (and likely youngest) $\gamma$-ray MSP to date, PSR~J1823$-$3021A in NGC~6624, which is also the first firm\footnote{\textit{AGILE} claimed a $4\sigma$ detection of PSR~J1824$-$2452 in M~28~\cite{Pellizoni09}. This was only later confirmed by {\it Fermi}~\cite{Johnson13}.} detection of a $\gamma$-ray MSP in a GC, allows constraints on the underlying mechanism of HE GC emission (Section~\ref{sec:Fermi}). Using the measured $\gamma$-ray light curve, one may define an `off-pulse window' -- a range in the cyclic normalized phase coordinate where the pulse is believed to be `switched off' ($0.07 < \phi < 0.60$ and $0.67 < \phi < 0.90$ in this case). Selecting off-pulse $\gamma$-rays only, one can then constrain the level of emission from other sources (e.g., other GC MSPs) or from the background. This technique is also used when searching for steady emission from putative pulsar wind nebulae surrounding younger pulsars~\cite{Ackermann11_PWN}. \textit{Fermi} did not detect any GeV point sources in the off-pulse window of PSR~J1823$-$3021A. A previous estimate~\cite{Abdo10,Tam11} assigned $\sim100$ $\gamma$-ray MSPs to this cluster. %assuming typical MSP spectral properties. 
We now know that this extraordinarily powerful MSP accounts for almost all of NGC~6624's HE emission and the off-pulse flux upper limit constrains the number of additional `ordinary' GC MSPs to $<32$. This implies that, at least for NGC~6624, the HE emission must be predominantly due to pulsed emission, believed to be magnetospheric curvature radiation (CR; Section~\ref{sec:CR}). Any unpulsed emission (e.g., due to inverse Compton (IC) scattering) is severely constrained in this case (Section~\ref{sec:Cheng}). Although this supports a pulsed origin for HE emission in other GCs, one can only speculate as to the general validity, given the plausible differences in MSP population properties and environments between GCs. %(i.e., the relative contribution of pulsed and unpulsed emission is generally unknown).

\section{Successes and challenges of the $\gamma$-ray MSP population model}
\subsection{Pulsed HE emission}
\label{sec:CR}
The fact that \textit{Fermi} has detected several GCs exhibiting spectra that are very reminiscent of pulsar spectra (hard power law with exponential cutoff at a few GeV) leads to the notion that a population of MSPs hosted within the GC may be cumulatively responsible for this HE emission~\cite{Abdo10}, strengthened by the discovery of $\gamma$-rays from PSR~J1823$-$3021A (see Section~\ref{sec:1823}). Indeed, this had been predicted~\cite{HUM05,Venter08,Venter09_GC} shortly before the discovery of 47~Tucanae in the HE band~\cite{Abdo09_Tuc}. The pulsed emission arises from particles being accelerated by electric fields inside the MSP magnetospheres, and suffering CR losses as they move along curved magnetic field lines, before being ejected beyond the light cylinder. Predictions for 47~Tucanae were reasonably close to the measured spectrum, and implied a population size of $\sim50$ members~\cite{Venter11_Fermi}, but overpredicted the spectral cutoff by a factor $\sim2$, suggesting some revision of the model (e.g., a lower electric field or smaller curvature radius may be needed; this may be attained by newer magnetospheric structures and different charge density properties). The predicted spectrum was also too hard~\cite{Venter09_GC}. It has to be borne in mind, however, that a pair-starved electric field has been assumed in these calculations. Subsequent light curve modelling of several $\gamma$-ray MSPs~\cite{Venter09} suggests that the bulk of this population may have screened electric fields, a possibility not considered previously within the standard assumption of MSP magnetospheres consisting of dipolar magnetic fields~\cite{HMZ02}.

\subsection{Unpulsed HE emission}
\label{sec:Cheng}
As an alternative to the CR mechanism, an IC scenario was considered to explain the HE fluxes seen by \textit{Fermi}~\cite{Cheng10}. The model solves a cosmic-ray diffusion equation (using a slightly different stellar photon energy density profile than~\cite{BS07}), and predicts that the bulk of the HE radiation comes from a region beyond the GC core (contrary to the findings of \cite{BS07}), so that GCs should be extended HE sources. The \textit{Fermi} flux may be reproduced for some combinations of model parameters in this scenario. This model also predicts spectral components that should be visible in the VHE domain in some cases. Such unpulsed IC components seem less dominant, however, in the case of the cluster NGC~6624, as explained in Section~\ref{sec:1823}. 

\subsection{Unpulsed VHE emission}
In addition to being sources of pulsed photons, 
%the large potential drops in MSP magnetospheres indicate that they are also sources of relativistic leptons: 
MSPs may also produce leptons with energies of up to a few TeV~\cite{Buesching08}. These leptons escape from the magnetospheres and may be reaccelerated in shocks formed by collisions of stellar winds in the cluster core~\cite{BS07,Bednarek11}. The leptons may then upscatter bright starlight and cosmic microwave background (CMB) photons to very high energies, leading to an unpulsed VHE spectral component, which gives an independent constraint\footnote{We are assuming that the number of visible HE MSPs $N_{\rm vis}\approx N_{\rm MSP}$.} (vs.\ pulsed emission) on $N_{\rm MSP}$ (depending on the cluster magnetic field and diffusion coefficient)~\cite{Venter09_GC}. It was found that 47~Tucanae and Terzan~5 may be visible for H.E.S.S., depending on model parameters~\cite{BS07,Venter09_GC}.
Using the H.E.S.S.\ upper limits on the VHE $\gamma$-ray emission from 47~Tucanae, we could infer a population of $\sim30-40$ MSPs, given a cluster magnetic field of $B\sim10\mu$G (but quite larger for $B<5\,\mu$G or $B>\,30\mu$G). 

\begin{figure}
  \begin{center}
  \includegraphics[width=9cm]{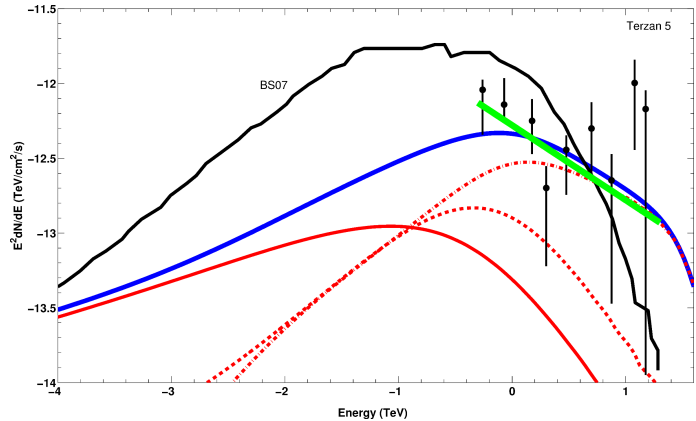}
  \end{center}
  \caption{\label{fig:spectrum} Predicted IC spectrum in the MSP scenario. The black solid line labelled BS07 is scaled from~\cite{BS07}, while the thick blue solid line represents the sum of the IC emission from three zones lying at increasing radii from the GC centre (indicated by the lower solid, dashed, and dashed-dotted red lines). We assumed a power-law lepton injection spectrum with minimum energy of 100~MeV, maximum energy of 100~TeV, photon index of 1.6, cluster field of 1~$\mu$\,G, and total power of $2\times10^{34}$ erg\,s$^{-1}$ (e.g., $N_{\rm MSP} =100$ if the conversion efficiency of spin-down power to particle power is 1\%). We also introduced a factor~3 scaling. This implies a larger value of $N_{\rm MSP}$, number of cluster stars ($N_{\rm star}>8\times10^5$), average stellar radius ($R>10^6$~cm), stellar temperature ($T>6~000$~K), or smaller GC distance ($d < 5.9$~kpc), or a combination of these~\cite{Prinsloo}. The green line and data points are from H.E.S.S.\ results~\cite{
Abramowski11_Ter5}.}
\end{figure}

We have now extended our calculations for Terzan~5 (Figure~\ref{fig:spectrum}), including a third large emission zone extending up to the tidal radius, a full calculation of the energy density profile~\cite{Prinsloo}, and assuming a power-law injection spectrum for the leptons. It has been expected that, since the stellar energy density profile is strongly peaked at the centre of the GC, the TeV flux should follow a similar profile, as the IC process depends on this target soft photon field~\cite{Domainko11}. However, even though the energy density drops steeply as one leaves the core region, the much larger size of the halo traps the particles for much longer, where they interact with the low-energy-density field. This increase in residence time outweighs the drop in soft photon energy density and emission from this region dominates the VHE spectrum in our model. %(dashed-dotted line in Figure~\ref{fig:spectrum}).

\subsection{Challenges}
Typical constraints derived from the measured GeV energy fluxes on $N_{\rm MSP}$ are not always so restrictive, given uncertainties in energy flux, GC distance, and average MSP beaming factors. This means that the level of the pulsed CR flux predictions is somewhat uncertain. Lack of evidence for spectral cutoffs around a few GeV for some GCs also challenge the cumulative CR interpretation for the HE GC emission~\cite{Bednarek12,Tam11}. Concerning VHE models, refinement of the soft photon energy density profile (e.g., adding a Galactic component~\cite{Cheng10}), cluster magnetic field profile, and particle transport calculations may be needed. The peculiar asymmetric, offset VHE source seen in coincidence with Terzan~5 also challenges a simplistic MSP scenario, but may be introduced by several factors, including a small MSP population at relatively large distances from the centre, formation of MSPs near the tidal radius in addition to the core~\cite{Tam11}, non-spherical photon target fields, proper motion of the GC, and an asymmetric diffusion coefficient~\cite{Cheng10}. Source extension, morphology, spectrum and energetics will continue to constrain GC models.

\section{Other sides of the VHE coin: alternative explanations}
\subsection{Explosive energy: a short $\gamma$-ray burst (GRB) relic in Terzan~5?}
Long GRBs ($>2$~s), signalling the death of massive stars~\cite{Gehrels05}, have been invoked as a possible origin for some of the many unidentified TeV sources observed by H.E.S.S.~\cite{Abramowski11_Ter5}. Similarly, the scenario of the VHE emission from Terzan~5 being due to the remnant of a short ($<2$~s), powerful GRB (thought to be due to a compact binary merger~\cite{Gehrels05}) has been put forward~\cite{Domainko11}. The main argument relies on the fact that the high-density environments of ancient GCs are conducive to the formation of compact binaries, and that the merger of the members (e.g., two neutron stars) may accelerate hadronic cosmic rays. The latter may interact with ambient target nuclei and decay into $\gamma$-rays via the $\pi^0$ channel. It was shown that for a target density of $n\sim0.1$~cm$^{-3}$, and a broken power-law cosmic-ray spectral shape with spectral index of 2.0 below 5~TeV, the total energy of hadrons of $E\approx10^{51}$~ergs may possibly be supplied by 
ultrarelativistic blast waves converting a significant part of the kinetic energy to cosmic ray particle energy. A break in the $\gamma$-ray spectrum in the GeV / TeV range would support such a scenario. The putative remnant age of $\sim10^4$ yr (estimated from the TeV source extension and depending on the hadron diffusion coefficient) corresponds roughly to the estimated rate of short bursts in the Milky Way ($\sim1$~event per $10^4$~yr, depending on the burst beaming factor) in the event of relatively slow diffusion. In this scenario, electrons accelerated by the blast wave scatter the stellar photons to produce diffuse non-thermal X-ray emission. Additional predicted signatures are faint thermal X-rays from hot thermal plasma in the remnant, faint nuclear line emission from radio-active decay of heavy nuclei ejected during the merger, and  faint ionization lines following from interaction of the GRB afterglow with the interstellar medium.

\subsection{Distant cousins: a partly WD origin?}
Instead of MSPs, a population of fast-rotating, magnetized WDs contained within GCs has been investigated as being responsible for (part of) the HE and VHE signals~\cite{Bednarek12}. An estimate involving the initial mass function implies that the number of all WDs in a GC may be as high as $\sim10^5$, dominating the MSP population by a factor of a few hundred. The evolutionary behaviour of the non-accreting WD energetics is furthermore very similar to the case of rotation-powered pulsars, although the typical surface magnetic fields may be lower ($\sim10^8$~G vs.\ $10^{8-9}$~G), rotational periods much longer ($\sim100$~s vs.\ $5$~ms), masses similar ($\sim0.8M_\odot\approx1.4M_\odot$), and radii larger ($\sim5\times10^8$~cm vs.\ $10^6$~cm). These parameters suggest that a single WD may have injected electrons with a power of up to $\sim10^{28}$~erg\,s$^{-1}$ (vs.\ $\sim10^{32}$~erg\,s$^{-1}$ for an MSP), but this would be less in the case of magnetic field decay. In addition, WDs created by WD-WD mergers in compact binary systems (having larger masses, smaller radii, and shorter periods than non-accreting WDs) may also contribute to the observed $\gamma$-ray flux, but at a much lower level (a factor $\sim10$). It has been shown~\cite{Bednarek12} that a few thousand WDs that have been created uniformly during the GC lifetime may produce a detectable VHE signal via upscattering of CMB and stellar soft photon fields (assuming mono-energetic electron injection spectra, and different models for WD formation, evolution, and magnetic field decay, and also depending on the electron acceleration process).

\section{Conclusion}
We have discussed GCs as multiwavelength objects, focusing on their $\gamma$-ray properties. Although some models  have proven reasonable, inevitable refinement is due in the wake of observations that are increasing both in quality and quantity.

\ack
This research is based upon work supported by the South African National Research Foundation.

\end{document}